\documentclass[conference]{IEEEtran}
\IEEEoverridecommandlockouts
\usepackage[normalem]{ulem}
\usepackage{commath}
\usepackage{cite}
\usepackage{amsmath,amssymb,amsfonts}
\usepackage{algorithmic}
\usepackage{graphicx}
\usepackage{subcaption}

\usepackage{epstopdf} 
\usepackage{textcomp}
\usepackage{xcolor}
\usepackage{float}
\usepackage{subcaption} 
\usepackage{dsfont}
\usepackage{enumitem}
\usepackage{commath}
\usepackage{dsfont}
\usepackage{amsthm}
\usepackage{svg}
\usepackage{epstopdf}
\usepackage{comment}

\usepackage[left=0.68in, right=0.68in,top=0.702in]{geometry}
\renewcommand{\baselinestretch}{0.95}

\theoremstyle{definition}

\setlist{noitemsep, topsep=0pt, partopsep=0pt, parsep=0pt}

\def\BibTeX{{\rm B\kern-.05em{\sc i\kern-.025em b}\kern-.08em
    T\kern-.1667em\lower.7ex\hbox{E}\kern-.125emX}}

\DeclareMathOperator*{\argmax}{argmax} 

\begin{document}
\bstctlcite{IEEEexample:BSTcontrol}
\title{
Sensing-Aided Beam Selection for UAV Networks
{}
}





\makeatletter
\def\IEEEauthorrefmark#1{\textsuperscript{%
  \ifcase#1
  \or *%
  \or \S%
  \or \P%
  \else \@ctrerr
  \fi}}
\makeatother

\author{%
\IEEEauthorblockN{%
Bora Bozkurt\IEEEauthorrefmark{1}\IEEEauthorrefmark{2},
Ali Gorcin\IEEEauthorrefmark{1}\IEEEauthorrefmark{2},
Cevdet Tosun\IEEEauthorrefmark{1}\IEEEauthorrefmark{2},\\
Mehmet Nuri Akıncı\IEEEauthorrefmark{2},
Ibrahim Hokelek\IEEEauthorrefmark{1},
Mehmet Kemal Özdemir\IEEEauthorrefmark{3}}
\IEEEauthorblockA{\IEEEauthorrefmark{1}%
Communications and Signal Processing Research (HİSAR) Lab., T{\"{U}}B{\.{I}}TAK B{\.{I}}LGEM, Kocaeli, Türkiye}
\IEEEauthorblockA{\IEEEauthorrefmark{2}%
Department of Electronics and Communication Engineering, Istanbul Technical University, {\.{I}}stanbul, Türkiye}
\IEEEauthorblockA{\IEEEauthorrefmark{3}%
Department of Electrical and Electronics Engineering, Medipol University, {\.{I}}stanbul, Türkiye}
}
\maketitle

\begin{abstract}
 Fast and low-overhead beam management is essential for the practical deployment of unmanned aerial vehicle (UAV) communication networks operating at millimeter-wave and higher frequencies. In this paper, we propose a radar-assisted beam selection framework for UAV networks that exploits spatial sensing information, including angle-of-departure and distance estimates, to efficiently identify  communication beams. Specifically, a digitally beamformed radar array with a relatively small number of antennas provides rapid, coarse UAV localization, while an analog-beamformed communication array, requiring fewer radio-frequency chains, exploits coarse localization to narrow the beam-search region and establish a communication link with the UAV. To provide theoretical insight into the expected overhead, we conduct a probabilistic analysis in which an approximation of the nearest-first beam selection overhead under idealized conditions is proposed and its statistics are derived for a Gaussian error model. The simulation results demonstrate that the proposed radar-assisted beam selection significantly reduces the number of beams that need to be examined, requiring on average at most $1.64$ candidate beams even under per-element radar phase errors of up to $\pm 100^\circ$.

 

\end{abstract}


\section{Introduction}
Integrated sensing and communication (ISAC) presents unprecedented opportunities in wireless networks as a key concept for enabling next generation applications and services \cite{qaisar2026role}. The promise of ISAC stems from significant gains such as reduced hardware costs, improved spectral efficiency, and lower energy consumption achieved through efficient integration of communication and sensing functionalities. 
The concept of ISAC that specifically addresses the problem of improving communication system performance using the sensing information is called sensing-assisted communication (SAC) \cite{mahmoud2025sensing}. The studies on SAC have mostly focused on optimizing beamforming using the sensing information for essentially improving the performance of millimeter-wave (mmWave) and sub-THz communications. 

In SAC systems with a dedicated sensing assistance, machine learning (ML) or more generally artificial intelligence (AI) has emerged as a widely adopted approach for beam management.
The survey by Brilhante et al. illustrates this landscape extensively \cite{brilhante2023literature}. It should be noted that recent studies have also explored fusion-based techniques for predicting the best beams from multi-modal data, for which an example is given in \cite{nazar2026enwar}. The need for such sophisticated AI-based techniques is due to the highly non-linear relationship between the optimal beams and the sensing data, induced by a channel with complex scattering conditions. 

Unlike terrestrial networks, non-terrestrial networks (NTNs) typically operate within a line-of-sight (LoS) dominated channel \cite{granelli2026survey}, which make structured and model-based beam management approaches particularly attractive. Hence, sensing-assisted beamforming and beam management studies for NTNs mostly follow analytical formulations. A comprehensive study by Yanpeng et al. \cite{yanpeng2024sensing} developed a beam management framework for mmWave unmanned aerial vehicle (UAV) networks that utilizes echoed communication signals alongside visual information at the base station (BS). The authors suggested a solution for user identification that matches the users' physical information with their temporary identifiers so that the beams can be allocated to their intended users. On the other hand, Benaya et al. \cite{benaya2025aerial} proposed a secure NTN architecture with a high-altitude platform station (HAPS) mounted ISAC BS and studied optimizing the sum spectral efficiency and secrecy rate of the system through beamforming and jamming. The scenario of sensing-assisted joint trajectory and beamforming optimization was investigated in \cite{zhou2025temporal, zhang2024sensing}. Finally, Chen et al. \cite{chen2026sensing} studied sensing-assisted beam tracking and beamwidth control suitable for UAV communications. 

Despite these advances, sensing-assisted beam management for UAV networks remains relatively underexplored, particularly for initial access and architectures that rely exclusively on radar for sensing assistance. To address this gap, we propose a radar-assisted beam selection framework for initial access in UAV-to-UAV communications. Unlike the approach in \cite{yanpeng2024sensing}, the proposed framework does not need an additional camera system such that a hierarchical beam selection strategy in which a radar array with a relatively small number of antennas uses digital beamforming to estimate the target direction in a single sweep, while a communication subarray with a larger number of antennas performs analog beam sweeping over the identified angular region. Another important feature is that a distance-aware early-stopping mechanism that uses the radar-estimated UE range to establish a received-power threshold, allowing the beam search to terminate once the received-power feedback indicates that a sufficiently strong beam has been found. These features jointly reduce beam-search overhead while maintaining narrow communication beams with limited RF-chain complexity. The main contributions of this paper are summarized as follows:

\begin{itemize}
\item We propose a radar-assisted beam selection framework for initial access in UAV-to-UAV communications. By combining digital beamforming at the radar array with localized analog beam sweeping at the communication array, the framework reduces the search space while limiting RF-chain requirements. In addition, a distance-aware early-stopping criterion exploits the radar-estimated UE range and received-power feedback to avoid unnecessary beam measurements.

\item We analyze the beam selection overhead probability under a nearest-first search strategy, which prioritizes candidate beams according to the angular distance between their steering directions and the radar-estimated UE angle of departure (AoD). Under simplifying assumptions, we derive an approximation of the search overhead and characterize its statistical behavior in the presence of Gaussian radar estimation errors.

\item The proposed framework is evaluated through extensive simulations, demonstrating that the radar-derived spatial and range information is instrumental to significantly reduce beam-search complexity while maintaining high-resolution beam
acquisition.

\end{itemize}

\begin{figure}
    \centering
    \includegraphics[width=1\linewidth]{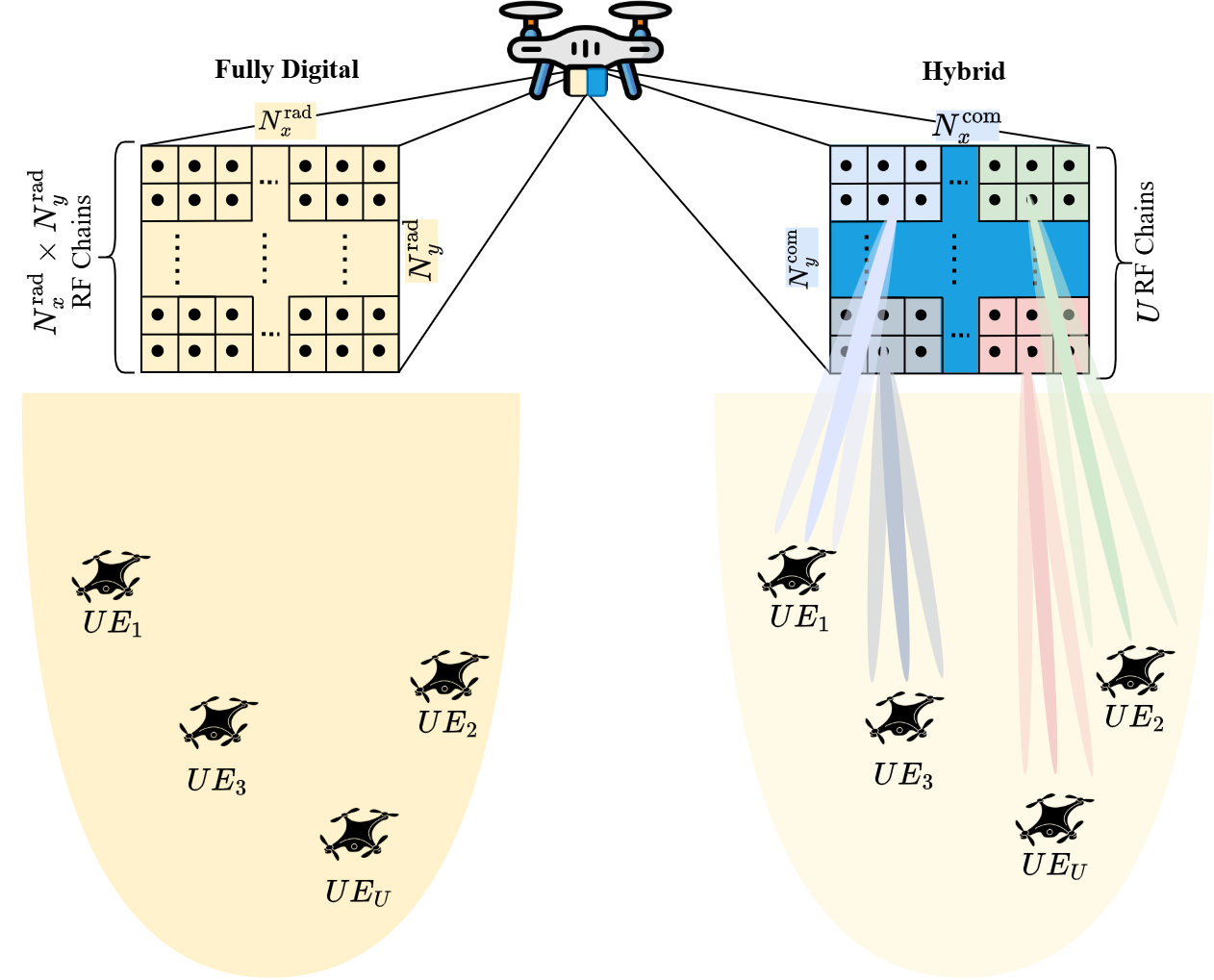}
    \caption{The radar-assisted beam selection framework.}
    \label{fig:system_model}\vspace{-10pt}
\end{figure}





\section{System Model}\label{sec:system_model}
We consider a sensing-assisted UAV network in which an aerial millimeter-wave (mmWave) base station (BS) serves $U$ aerial UEs. As illustrated in Fig. \ref{fig:system_model}, the BS employs two uniform rectangular array (URA) subarrays for radar sensing and communication, comprising $N^\mathrm{rad}_x \times N^\mathrm{rad}_y$ and $N^\mathrm{com}_x \times N^\mathrm{com}_y$ antenna elements, respectively. To reduce hardware complexity and cost, both functionalities can share a common hardware platform. The communication array is partitioned into $U$ equal-sized subarrays, each connected to a dedicated RF chain, enabling simultaneous service to all UEs. In contrast, the radar array employs a fully digital architecture, with a dedicated RF chain for each antenna element. 

The air-to-air (A2A) channel allows the radar to capture accurate spatial information about the users and other passive targets nearby. To describe the radar sensing model, let $\tilde{s}(t)$ represent the baseband sensing signal, which is upconverted to a carrier frequency $f_c$ for transmission. With the help of background removal, at the the radar's $n^\mathrm{th}$ antenna element ($n = 1, \dots, N^\mathrm{rad}_xN^\mathrm{rad}_y$), the signal reflected from the target is observed as
\begin{equation}\label{eq:recv_rad_sig}
x_n(t) \!= \! \mathrm{Re}\!\left\{\!\tilde{s}\left(t - \frac{2d+d_n}{c}\right)\!e^{j\left(2\pi f_c\left(t - \frac{2d+d_n}{c}\right)+\hat\phi_n(t)\right)}\!\right\} \!+\! z_n(t),
\end{equation}
where $d$ denotes the distance between the target and the radar's array, $d_n$ represents the $n^\mathrm{th}$ antenna element's incremental distance to the target relative to the array center, $c$ is the speed of light, $\hat{\phi}_n(t)$ denotes the phase noise, and $z_n(t)$ is the additive noise. From \eqref{eq:recv_rad_sig}, the baseband received signal obtained from the $n^\mathrm{th}$ antenna element becomes
\begin{equation}
\tilde{x}_n(t) = \tilde{s}\left(t - \frac{2d+d_n}{c}\right)e^{-j\left(2\pi f_c\left( \frac{2d+d_n}{c}\right) - \hat\phi_n(t)\right)} + \tilde{z}_n(t),
\end{equation}
where $\tilde{z}_n(t)$ represents additive noise. Since the transmitted sensing signal $s(t)$ is known, from the received baseband signals $\tilde{x}_n(t)$, the radar is able to derive the spatial information of the target by utilizing the appropriate methods. More specifically, we assume that the distance to the target and its angle of departure (AoD) relative to the BS are obtained via matched filtering and the Multiple Signal Classification (MUSIC) algorithm, respectively.

After detecting all potential UEs using the radar and obtaining their spatial information, the BS performs a codebook-based beam selection procedure for each UE using the communication subarray dedicated to that UE. Since the beam selection is carried out separately at UEs and BSs in practical systems \cite{ziao2025review}, this study considers the BS side beam selection only and assumes that there is a single antenna at UEs. Specifically, the BS selects a beam for each UE by observing the beam index feedback, which is determined by the UE from the received signal strength (RSS) values. 

Each communication URA subarray comprises $N_x \times N_y$ antenna elements, with half-wavelength spacing along both array axes. Accordingly, each subarray contains $N_{A} = N_xN_y$ antenna elements, and the total number of communication antenna elements across all $U$ users is $N_AU = N_x^\mathrm{com}N_y^\mathrm{com}$. Define the normalized symbol transmitted by the BS to the $u^\mathrm{th}$ UE with $s_u \in \mathbb{C}$, $\abs{s_u} = 1$. At time slot $t$ ($t = 1, \dots, T$), the symbol is multiplied by a weight vector, i.e., beam, $\mathbf{f}_u[t] \in \mathcal{F}$, belonging to the codebook $\mathcal{F}$. We consider an oversampled 2D discrete Fourier Transform (DFT) codebook inspired by the spatial Fourier structure
used in NR codebook designs \cite{ziao2025review}, with integer oversampling factor $O \geq 1$. Prior to defining the codebook, we introduce the normalized DFT vector
\begin{equation}\label{eq:dft_vec}
\mathbf{w}_N(k) = \frac{1}{\sqrt{N}}\left[1, e^{-j \pi \mu_N(k)}, \dots, e^{-j \pi (N-1)\mu_N(k)}\right]^T,
\end{equation}
where $\mu_N(k) = 2 k / (ON) - 1$ represents the sine domain steering angle. Using \eqref{eq:dft_vec} and the Kronecker product operator $\otimes$, the codebook is written as
\begin{equation}
\begin{aligned}
& \mathcal{F}  = \{\mathbf{w}(k,m) =\mathbf{w}_{N_x}(k)\otimes \mathbf{w}_{N_y}(m):  \\ 
& k \in \{1,\dots, ON_x \} \text{ and } m \in \{1, \dots, 
ON_y \} \}.
\end{aligned}
\end{equation}
After multiplication with the weight vector, the signal is transmitted through the channel and is received at the $u^\mathrm{th}$ UE as
\begin{equation}\label{eq:recv_sig}
y_u[t] = \sqrt{P}\mathbf{h}_{u,u}^T\mathbf{f}_u[t]s_u + \sqrt{P}\sum_{i \neq u}\mathbf{h}_{i,u}^T\mathbf{f}_i[t]s_i + z_u[t],
\end{equation}
where $P$ is the per-subarray transmission power, $\mathbf{h}_{i,u} \in \mathbb{C}^{N_A \times 1}$ denotes the narrowband flat fading channel between the $i^\mathrm{th}$ subarray and $u^\mathrm{th}$ UE, which is assumed to be constant during the beam selection period, and $z_u[t] \sim \mathcal{CN}(0, \sigma^2)$ is the additive white Gaussian noise distributed independently with respect to (w.r.t.) $t$ and $u$. Accounting for the summation term in \eqref{eq:recv_sig} that represents the interference, the instantaneous achievable rate of the $u^\mathrm{th}$ UE at time slot $t$ is given by
\begin{equation}\label{eq:achv_rate}
R_u[t] = \log_2\left( 1 + \frac{\abs{\mathbf{h}_{u,u}^T\mathbf{f}_u[t]}^2}{\abs{\sum_{i \neq u}\mathbf{h}_{i,u}^T\mathbf{f}_i[t]s_i}^2 +\sigma^2/P}\right).
\end{equation}

We model the narrowband flat fading A2A channel \cite{channelModelA2A} between the $i^\mathrm{th}$ subarray and $u^\mathrm{th}$ UE with
\begin{equation}\label{eq:aav_chan}
   \mathbf{h}_{i,u}\!= \!\sqrt{\frac{K}{K+1}}\mathbf{h}_{i,u}^\mathrm{LoS}\!+\!\sqrt{\frac{\eta^\mathrm{RSR}}{K+1}}\mathbf{h}_{i,u}^\mathrm{RSR} \! + \!\sqrt{\frac{\eta^\mathrm{NL}}{K+1}}\sum_{\ell=2}^{L-1}\mathbf{h}_{i,u}^\ell.
\end{equation} 
where, $\mathbf{h}_{i,u}^\mathrm{LoS} \in \mathbb{C}^{N_A \times 1}$ represents the line of sight (LoS) channel, $\mathbf{h}_{i,u}^\mathrm{RSR}\in \mathbb{C}^{N_A \times 1}$ denotes the rooftop specular reflection (RSR), $\mathbf{h}_{i,u}^\ell \in \mathbb{C}^{N_A \times 1}$ represents the non-line-of-sight (NLoS) components and $K$ is the Rician K-factor. The power correction factors of the RSR and NLoS channels are given by $\eta^\mathrm{RSR}$ and $\eta^\mathrm{NL}$, respectively. Expanding the individual channel terms in \eqref{eq:aav_chan}, we obtain
\begin{equation}\label{eq:aav_chan_expanded}
\begin{aligned} 
    &\mathbf{h}_{i,u}\!=\!{\sqrt{\frac{K}{K+1}}\beta_{i,u}\mathbf{a}(\phi_{i,u},\theta_{i,u})} \!\! + \!\! {\sqrt{\frac{\eta^\mathrm{RSR}}{K+1}}\beta_{i,u}^\mathrm{RSR}\!\mathbf{a}(\phi_{i,u}^\mathrm{RSR}\!\!,\theta_{i,u}^\mathrm{RSR})} \\
    &+{\sqrt{\frac{\eta^\mathrm{NL}}{K+1}}\sum_{\ell=2}^{L-1}\beta_{i,u}^\ell\mathbf{a}(\phi_{i,u}^\ell,\theta_{i,u}^\ell)},
\end{aligned}
\end{equation} 
where \begin{equation}
\begin{aligned}
&\mathbf{a}(\phi, \theta) \! = \!\! \frac{1}{\sqrt{N_A}} \! \left[1, e^{j\pi \cos(\phi) \sin(\theta)}, \dots, e^{j(N_x-1)\pi \cos(\phi) \sin(\theta)}\right]^T \\ 
& \otimes \left[1, e^{j\pi \sin(\phi) \sin(\theta)}, \dots, e^{j(N_y-1)\pi \sin(\phi) \sin(\theta)}\right]^T 
\end{aligned}
\end{equation}
denotes the azimuth-elevation URA steering vector for half wavelength spacing and $\beta_{i,u}$, $\beta^\mathrm{RSR}_{i,u}$, and $\beta^\ell_{i,u}$ represent the path coefficients. Finally, for convenience, we also define the mapping $\psi$ from $[-\pi/2,\pi/2]^2$ into the unit disk $\mathbb{D} \subset \mathbb{R}^2$ that allows conversion between sine and angle domains as 
\begin{equation}
\psi= (\phi, \theta) \mapsto (\cos(\phi) \sin(\theta), \sin(\phi) \sin(\theta)).
\end{equation}

\section{Radar-assisted Beam Selection Framework}\label{subsec:radar_assisted_bt}

As illustrated in Fig. \ref{fig:system_model}, the proposed radar-assisted beam selection framework aims to eliminate the initial communication beam-sweeping stage by exploiting the wider beams formed by the radar array to obtain coarse spatial information. This enables direct beam refinement using narrower DFT beams. Specifically, for each UE, the BS employs its radar to obtain coarse estimates of the dominant AoD and the corresponding UE's distance. As in \cite{yanpeng2024sensing}, we assume that the BS is able to determine which spatial information (AoD and distance) belongs to which UE by pairing the UEs' spatial information with their temporary identifiers assigned by the network. In this paper, we do not consider this identity matching problem and refer the interested readers to \cite{yanpeng2024sensing}. 

Recall that the spatial information provided by the radar for the $u^\mathrm{th}$ UE consists of the estimated (dominant LoS path) AoD $(\hat\phi_{u,u}, \hat\theta_{u,u})$, and the estimated UE distance $\hat{d}_u$. Due to noise and other imperfections in the radar's estimation process, these estimates are subject to errors. Utilizing the estimated AoD of the $u^\mathrm{th}$ UE, $(\hat\phi_{u,u}, \hat\theta_{u,u})$, the BS performs a beam sweep over the $u^\mathrm{th}$ communication subarray using a subset of the codebook $\mathcal{F}$ to identify the beam that maximizes the received power. Specifically, we define $\mathcal{C} \subset \mathcal{F}$ as the subset of codebook elements whose steering angles in the sine domain lie within the rectangular region $D =[\hat\varphi_\mathrm{min}, \hat\varphi_\mathrm{max}] \times [\hat\vartheta_\mathrm{min}, \hat\vartheta_\mathrm{max}]$. This rectangle represents the coarse angular search area identified by the radar, centered at the estimated sine-domain AoD $(\hat\varphi_{u,u}, \hat\vartheta_{u,u})=\psi(\hat\phi_{u,u}, \hat\theta_{u,u})$. Consequently, the codebook subset can be expressed as
\begin{equation}
\mathcal{C} = \left\{\mathbf{w}(k,m) \in \mathcal{F} : (\mu_{N_x}(k), \mu_{N_y}(m)) \in D \right\}.\label{eq:codebook_subset}
\end{equation}

In addition to the AoD estimates, the BS exploits the estimated UE distance, $\hat{d}_u$, to further reduce beam selection overhead. Specifically, the distance estimate is used to determine an approximate received-power threshold, $P_\mathrm{th}$, that serves as an early stopping criterion for the beam sweep. In conventional beam selection, the UE evaluates all candidate beams and feeds back the index of the beam that maximizes the received power. In contrast, the proposed scheme terminates the beam sweep as soon as a candidate beam achieves a received power greater than or equal to \(P_{\mathrm{th}}\). To this end, the BS transmits the pre-determined $P_\mathrm{th}$ value to the UE before initiating beam selection. As the beams are swept, the beam that triggers this condition is then selected for the corresponding UE. 

For notational simplicity, we focus on a single UE and its corresponding communication subarray, omitting the UE index $u$ and the subarray index $i$ henceforth. Accordingly, we write $\mathbf{f}[t] = \mathbf{f}_u[t]$, $\mathbf{h} = \mathbf{h}_{i,u}$ (for $i=u$), and similarly for other indexed quantities. Let the time index $t$ denote an arbitrary ordering of the codebook elements as 
\begin{equation}\label{eq:enumeration}
\mathbf{f}[t]  = \mathbf{w}(k[t], m[t]), \quad t =1, \dots, T,
\end{equation}
where $T = \abs{\mathcal{C}}
$ represents the number of elements in the set $\mathcal{C}$. Assuming well-separated UEs and thus negligible interference, the optimal beam index maximizing the considered UE's received power is
\begin{equation}
\tau_\mathrm{opt}  =  \argmax_{1\leq t \leq T} \abs{\sqrt{P}\mathbf{h}^T\mathbf{f}[t]}^2,
\end{equation}
On the other hand, according to the stopping condition previously described, the BS selects the beam index 
\begin{equation}\label{eq:tau_bs}
\tau = \min \left\{t \in \{1, \dots, T \}:  \abs{\sqrt{P}\mathbf{h}^T\mathbf{f}[t] + z[t]}^2 \geq P_\mathrm{th} \right\}.
\end{equation}
If the set on the right hand side of \eqref{eq:tau_bs} is empty, then the BS simply selects the index of the beam maximizing the received power. A natural choice for the threshold is $P_\mathrm{th}=\alpha P \abs{\beta}^2 $, where $\alpha \in (0,1)$ is a tunable constant and $P\abs{\beta}^2$ represents the maximum achievable noiseless received power under a single-path channel, neglecting interference. Note that owing to the dominant LoS connectivity, the path loss factor $\abs{\beta}$ can be approximated using the UE distance estimated by the BS and the free-space path loss formula.

\section{Analysis of Nearest-First Beam Selection Overhead Probability}\label{sec:overhead_analysis}

This section analyzes the beam selection overhead probability associated with a distance-based search strategy, referred to as the nearest-first selection overhead (NFSO). In this strategy, candidate beams are examined in ascending order of the angular distance between their steering directions and the radar-estimated AoD, following a center-outward search pattern. To facilitate a tractable analysis of the proposed early-stopping mechanism, we first introduce a set of simplifying assumptions. Specifically, we consider a high-SNR regime in which noise and interference are neglected, and assume a single-path channel model given by $\mathbf{h} = \beta  \mathbf{a}(\phi,\theta)$. 
For analytical tractability, we assume that the path-loss coefficient $\abs{\beta}^2$ is perfectly estimated from the UE's distance. In practice, however, such estimates are subject to errors. The impact of these imperfections is evaluated in the subsequent simulation section by modeling phase errors at individual antenna elements as independent uniformly distributed random variables.

Secondly, we set $P_\mathrm{th} = P\alpha\abs{\beta}^2$ and assume that there exists a unique optimal beam $\mathbf{f}[t] \in \mathcal{C}$ whose received power satisfies $\lvert\sqrt{P}\mathbf{h}^T\mathbf{f}[t]\rvert^2 \geq P_\mathrm{th}$. 
Recall that $(\hat\varphi, \hat\vartheta) = \psi(\hat\phi,\hat\theta)$ and $(\varphi, \vartheta) = \psi(\phi,\theta)$ denote the sine domain AoD estimated by the radar and the actual sine domain AoD, respectively. These assumptions imply that 
\begin{equation}\label{eq:tau_optimality}
\begin{aligned}  
\tau = \tau_\mathrm{opt}  \iff (\varphi, \vartheta) \in D[\tau]
\end{aligned}
\end{equation}
where 
\begin{equation}\label{eq:rectangle_t}
\begin{aligned}  
D[t] = &\left[\mu_{N_x}(k[t]) - \frac{1}{ON_x}, \mu_{N_x}(k[t]) + \frac{1}{ON_x} \right] \\
    & \times \left[\mu_{N_y}(m[t]) - \frac{1}{ON_y}, \mu_{N_y}(m[t]) + \frac{1}{ON_y} \right].
\end{aligned}
\end{equation}
Note that the equivalence in \eqref{eq:tau_optimality} follows from the uniformly spaced structure of the DFT beams.
The relation $\tau = \tau_\mathrm{opt}$ enables us to neglect beam misalignment, simplifying the analysis. In this scenario, the selection overhead becomes purely influenced by the radar's estimation error and the order of the codebook beams. For a more general treatment involving misalignment probability, readers are referred to \cite{liu2017millimeter}. 
Finally, we note that with these assumptions, it is ensured that the true AoD always belongs to the region detected by the radar, i.e., $(\varphi,\vartheta) \in D \subset \bigcup_{t=1}^TD[t]$. This is reasonable given that $D$ is large and the estimation error is typically small.

    \subsection{NFSO Approximation for Naive Distance Based Search}
\begin{figure}[t]
    \centering
    \includegraphics[width=0.73\linewidth]{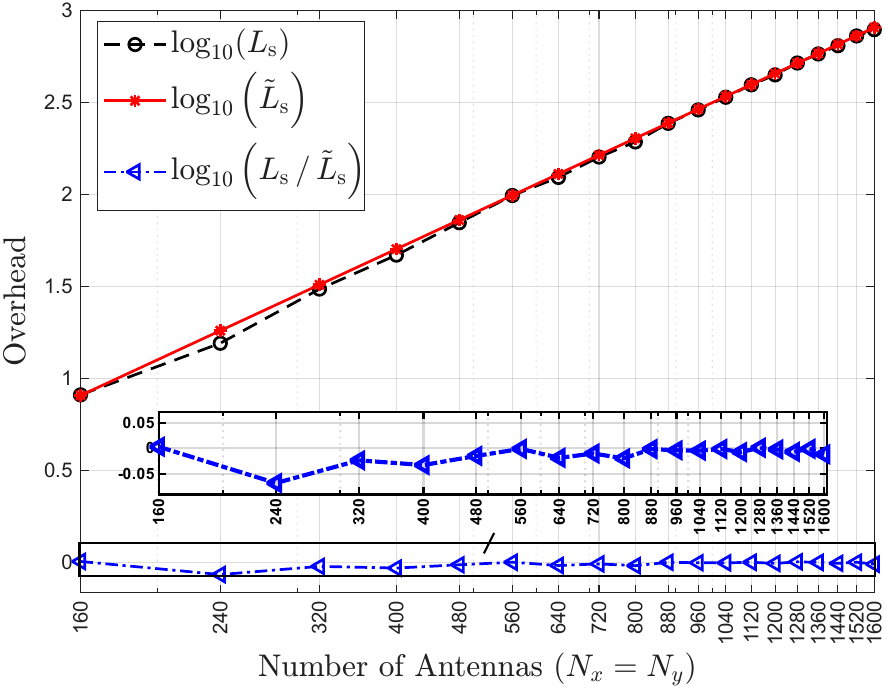}
    \caption{Values of $L_\mathrm{s}$ and $\tilde{L}_\mathrm{s}$ w.r.t. $N_x = N_y$.}
    \label{fig:approx_vs_actual}\vspace{-10pt}
\end{figure}

We now derive an approximation for the NFSO under the aforementioned assumptions. We model the total Euclidean error of the estimated AoD as
\begin{equation}
\delta = \sqrt{(\hat\varphi - \varphi)^2 + (\hat\vartheta - \vartheta)^2}= \sqrt{\delta_x^2 + \delta_y^2}, 
\end{equation} 
where the random variables $\delta_x$ and $\delta_y$ represent the estimation error along the x and y axes, respectively. 
Various efficient closed-loop techniques have been proposed for estimating the true AoD and, hence the best beam, by utilizing the received power values. 
However, for the sake of tractability, here we adopt a naive nearest-first distance based search strategy, which can be seen as a conventional benchmark. According to this search strategy, the beams are ordered such that the distance between the sine domain steering angle of $\mathbf{f}[t]$ and the estimated AoD $(\hat\varphi, \hat\vartheta)$ is non-decreasing w.r.t. $t$. As can be observed from \eqref{eq:tau_optimality}, this implies that a search over the nearest rectangles, $D[1], D[2], \dots$, around the estimated AoD is conducted, until the condition $\lvert\sqrt{P}\mathbf{h}^T\mathbf{f}[t]\rvert^2 \geq P_\mathrm{th} \iff t =\tau = L_\mathrm{s} \iff (\varphi, \vartheta) \in D[L_\mathrm{s}]$ is met. Here, $L_\mathrm{s}$ denotes the NFSO. Let $\mathcal{D} = \bigcup_{t=1}^{L_\mathrm{s}} D[t]$. Since all rectangles $D[t]$ have equal area, we have
\begin{equation}\label{eq:L_a}
L_\mathrm{s} = \frac{\mathrm{A}(\mathcal{D})}{\mathrm{A}(D[1])} = \frac{O^2N_xN_y\mathrm{A}(\mathcal{D})}{4},
\end{equation}
where $\mathrm{A}$ is the set function that yields the total area of a rectangle in $\mathbb{R}^2$. In our scenario, the DFT beams used for communication have a small beam width. Therefore, we consider the case where $ON_x$ and $ON_y$ are large enough so that $r = \sqrt{(2/(ON_x))^2 + (2/(ON_y))^2} < \delta$. For a given $\delta$, it is clear that $B((\hat\varphi, \hat\vartheta), \delta-r) \subset \mathcal{D} \subset B((\hat\varphi, \hat\vartheta), \delta+r)$, where $B((\hat\varphi, \hat\vartheta), \delta \pm r) \subset \mathbb{R}^2$ is the ball with center $(\hat\varphi, \hat\vartheta)$ and radius $\delta \pm r$. Hence,
\begin{equation}\label{eq:L_a_ulb}
\begin{aligned}
& \mathrm{A}(B((\hat\varphi, \hat\vartheta), \delta-r)) \leq \mathrm{A(\mathcal{D}}) \leq \mathrm{A}(B((\hat\varphi, \hat\vartheta), \delta+r)) \\
& \Longleftrightarrow \pi(\delta-r)^2 \leq \mathrm{A}(\mathcal{D}) \leq \pi(\delta + r)^2 \\
& \Longleftrightarrow \frac{O^2N_xN_y\pi(\delta-r)^2}{4} \leq L_\mathrm{s} \leq \frac{O^2N_xN_y\pi(\delta + r)^2}{4}
\end{aligned}
\end{equation}
from the monotonicity of the set function $\mathrm{A}$ and \eqref{eq:L_a}.
Furthermore, as $N_x,N_y\rightarrow\infty$, we get $r\rightarrow0$, implying
\begin{equation}\label{eq:L_a_lim}
\lim_{N_x,N_y \to \infty}\frac{4L_\mathrm{s}}{O^2N_xN_y \pi \delta^2} =1
\end{equation}
from \eqref{eq:L_a_ulb}. Finally, using \eqref{eq:L_a_lim} and the bounds in \eqref{eq:L_a_ulb}, we obtain the suggested NFSO approximation as
\begin{equation}\label{eq:L_a_approx}
 L_\mathrm{s} \approx \tilde{L}_\mathrm{s} = \frac{O^2N_xN_y \pi \delta^2}{4}.   
\end{equation} 
To illustrate the accuracy of the approximation, Fig. \ref{fig:approx_vs_actual} shows the numerical results of the actual NFSO $L_\mathrm{s}$ and the approximation $\tilde{L}_\mathrm{s}$ w.r.t. the number of antennas. For these results, we set $N_x=N_y$, $O=1$, $(\hat\varphi, \hat\vartheta) = \psi({\pi}/{3},{\pi}/{3})$, and distribute $(\varphi, \vartheta)$ uniformly on a circle such that the Euclidean distance between the actual AoD $(\varphi, \vartheta)$ and the estimated AoD $(\hat\varphi, \hat\vartheta)$ is equal to $\delta$. 
In addition, we fix $\delta =0.02$, thus $r < \delta$ for all array configurations. Fig. \ref{fig:approx_vs_actual} shows that the approximation $\tilde{L}_\mathrm{s}$ follows $L_\mathrm{s}$ accurately over a wide range of antenna numbers.  

\subsection{Statistics of the NFSO Approximation for Gaussian Error} 
Consider independent zero-mean Gaussian errors along both axes, $\delta_x \sim \mathcal{N}(0,\varsigma^2_x)$ and $\delta_y \sim \mathcal{N}(0,\varsigma^2_y)$. Depending on the magnitude of the radar error relative to the rectangle size $2/(ON_x) \times 2/(ON_y)$, two regimes arise.

\textit{Large errors:} When $r < \delta$, \eqref{eq:L_a_approx} applies and $\delta^2$ follows a squared Hoyt distribution \cite{romero2015on}, which reduces to an exponential distribution for $\varsigma_x = \varsigma_y$. Consequently, $\mathrm{E}[\tilde{L}_\mathrm{s}] = \frac{O^2N_xN_y\pi(\varsigma_x^2 + \varsigma_y^2)}{4}$, i.e., the NFSO grows linearly with both the radar error variance and the number of communication antennas, and for $\varsigma_x = \varsigma_y$, $\tilde{L}_\mathrm{s}$ is exponentially distributed with a standard deviation equal to its mean.

\textit{Small errors:} In practice, the radar error is often much smaller than a single rectangle, i.e., $\varsigma_x \ll 1/(ON_x)$ and $\varsigma_y \ll 1/(ON_y)$. In this regime, the NFSO is governed by the position of the true AoD within $D[\tau_\mathrm{opt}]$ rather than by $\delta$ alone. Place the origin at the corner of $D[\tau_\mathrm{opt}]$ nearest to the true AoD, and let $(a_x, a_y)$, with $a_x \in [0, 1/(ON_x)]$ and $a_y \in [0, 1/(ON_y)]$, be the coordinates of the true AoD. The estimate lies with high probability in one of the four rectangles sharing this corner, whose centers are $\left(s_x/(ON_x), s_y/(ON_y)\right)$ with $s_x, s_y \in \{\pm 1\}$, where $s_x = s_y = +1$ corresponds to $D[\tau_\mathrm{opt}]$. Denoting the estimate by $(X, Y) = (\delta_x + a_x, \delta_y + a_y)$, its squared distance to the center indexed by $(s_x, s_y)$ is $c - \frac{2}{O}\left(s_xX/N_x + s_yY/N_y\right)$, where $c = X^2 + Y^2 + 1/(ON_x)^2 + 1/(ON_y)^2$. Hence, the nearest-first order follows the score $s_xX/N_x + s_yY/N_y$. Comparing the score of $D[\tau_\mathrm{opt}]$, i.e., $X/N_x + Y/N_y$, with those of the rectangles indexed by $(-1,+1)$, $(+1,-1)$, and $(-1,-1)$ shows that they precede $D[\tau_\mathrm{opt}]$ if and only if $X < 0$, $Y < 0$, and $X/N_x + Y/N_y < 0$, respectively, which yields
\begin{equation}\label{eq:L_corner}
\tilde{L}_\mathrm{s} = 1 + \mathds{1}\{X<0\} + \mathds{1}\{Y<0\} + \mathds{1}\left\{\frac{X}{N_x} + \frac{Y}{N_y} < 0\right\},
\end{equation}
where $\mathds{1}\{\cdot\}$ is the indicator function and $X,Y$ are Gauss r.v.s. Taking the expectation of \eqref{eq:L_corner} gives
\begin{equation}\label{eq:EL_corner}
\begin{aligned}
\mathrm{E}[\tilde{L}_\mathrm{s}] \approx {} & 1 + Q\!\left(\frac{a_x}{\varsigma_x}\right) + Q\!\left(\frac{a_y}{\varsigma_y}\right) + Q\!\left(\frac{a_x/N_x + a_y/N_y}{\sqrt{\varsigma_x^2/N_x^2 + \varsigma_y^2/N_y^2}}\right)
\end{aligned}
\end{equation}
where $Q(\cdot)$ is the Gaussian tail function. Unlike large error case, \eqref{eq:EL_corner} accounts for the position of the true AoD within $D[\tau_\mathrm{opt}]$.

\section{Simulation Results} 

\label{sec:sim_results}
The simulation environment consists of a radar-assisted BS at 500~m altitude and four UEs with $-10$~dBsm radar cross section, located at $(x,y,z)={(80,0,520),(40,60,500),(60,-30,480),(100,-60,540)}$~m. The radar employs a 24~GHz linear frequency-modulated (LFM) waveform with 200~MHz bandwidth and 50~$\mu$s pulse duration, an $8\times8$ URA and MUSIC-based DoA estimation with 64 snapshots. Per-element phase errors of $\mathcal{U}(-\Delta, \Delta)$, with $\Delta \in [0^\circ, 100^\circ]$, model array calibration errors, and AWGN is added at radar SNRs of $-10$, $0$, and $20$~dB. The communication subsystem operates at 28~GHz using a $32\times32$ URA divided into four independently controlled $16\times16$ subarrays with Chebyshev tapering to suppress sidelobes and $O=1$ DFT beams. Propagation follows the Rician-based air-to-air geometric stochastic channel described in Section~\ref{sec:system_model} and \cite{channelModelA2A}, with $K=15$~dB, $\eta^\mathrm{RSR}=0.4$, $\eta^\mathrm{NL}=0.6$ and $L=10$ paths.

The beam alignment procedure exploits coarse UE location estimates obtained from the $8\times8$ radar array under imperfect channel. Since the $3$~dB beamwidth is inversely proportional to the number of antenna elements along a given dimension, the $16\times16$ communication array provides a beamwidth approximately four times narrower than that of the radar array. Accordingly, the region around each radar-derived UE location is discretized into a $2\times2$ grid, yielding a codebook of ($\abs{\mathcal{C}}=4$ in \eqref{eq:codebook_subset}) candidate communication beams. The beams are searched in a center-outward order, starting from the candidate closest to the estimated UE location and progressively exploring the surrounding candidates. The search terminates when the received signal strength (RSS) exceeds the predefined threshold, thereby trading residual beam alignment loss for reduced search latency. The threshold is set 5~dB below the maximum-realizable received power assuming accurate range estimation, corresponding to $\alpha=10^{-5/10}\approx0.316$. The impact of range-estimation errors on threshold calibration is left for future investigation.

\begin{figure}[t]
    \centering  \includegraphics[width=0.75\linewidth]{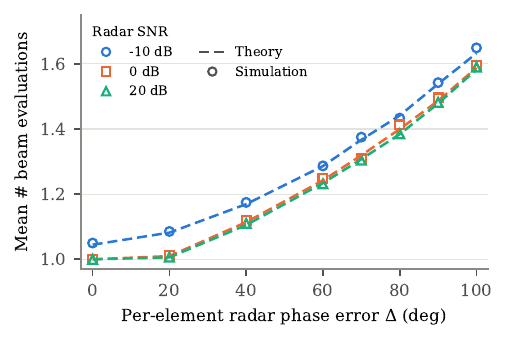} 
    \caption{The mean number of beam evaluations vs radar phase error $\Delta$ (markers: simulation and dashed lines: theory~\eqref{eq:EL_corner}).}
 \label{fig:3d_hist}\vspace{-12pt}
\end{figure}

Figure~\ref{fig:3d_hist} shows the mean number of beam evaluations for UE-3. Since the beams are evaluated center-outward, the selected index equals the number of evaluations. The theoretical curves follow from \eqref{eq:EL_corner} with the radar-measured error variances, which satisfy the small-error condition and match the simulation within $1\%$. Even at $\Delta = 100^\circ$, only $1.5$ evaluations are required for all SNRs. Without radar assistance, exhaustive measurement of all $256$ codebook beams would be required. Even with radar assistance, without the proposed threshold, all $16$ candidate beams would need to be measured. Both cases require more measurements and longer beam-selection time than the proposed approach. 

\begin{figure}[!h]
   \centering
    \includegraphics[width=0.47\textwidth]{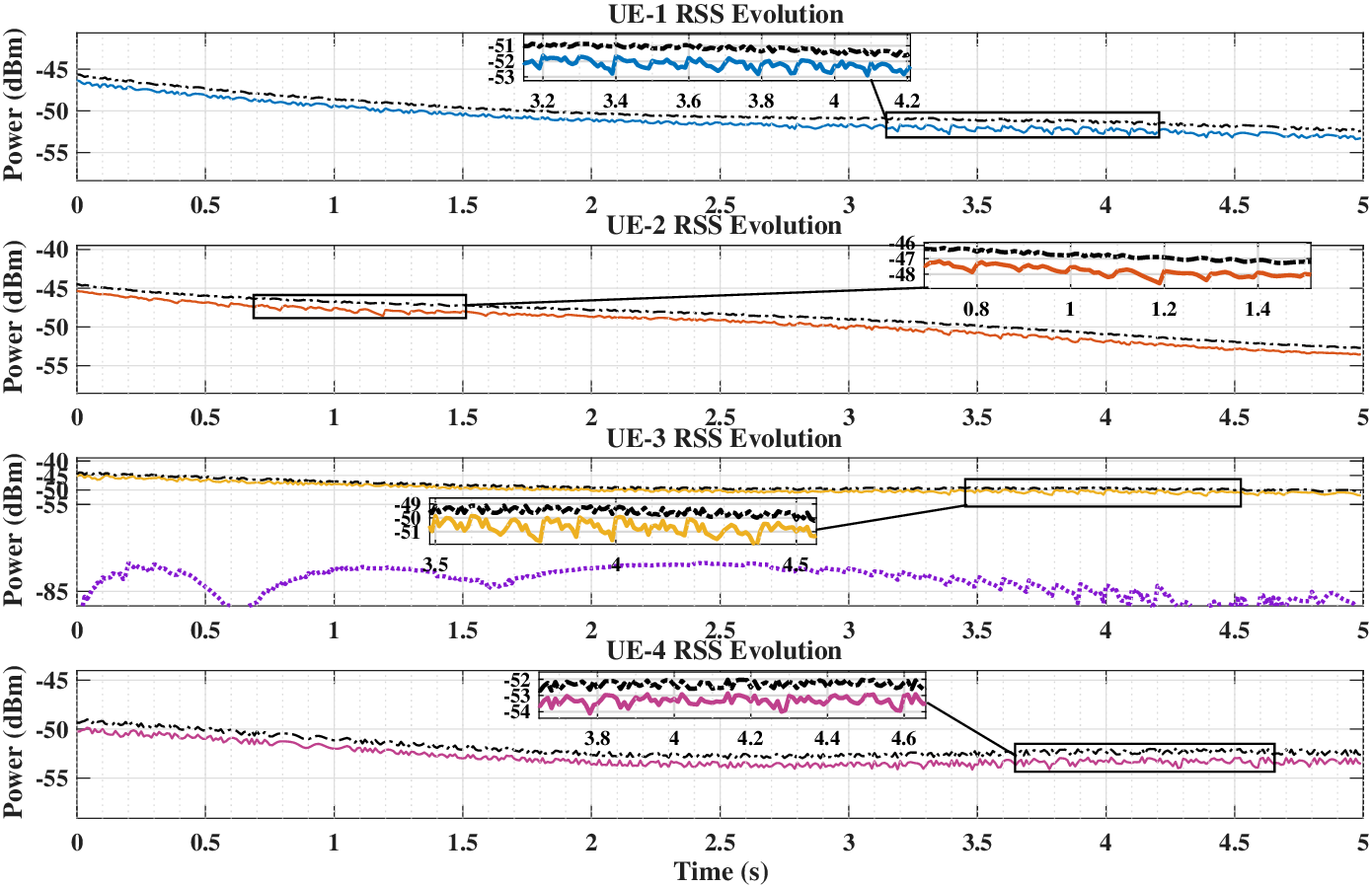} 
    \caption{RSS performance of the four UEs, where the black dashed curves denote the maximum achievable RSS with the best beam alignment, the colored solid curves show the RSS achieved by the proposed beam selection while the purple dotted curve shows the RSS of the intentionally nulled UE-3.} 
    \label{fig:3ue_tracking}\vspace{-17pt}
\end{figure}

Figure~\ref{fig:3ue_tracking} illustrates the RSS evolution of the four UEs. The black dashed curves indicate the maximum achievable RSS with ideal beam alignment, while the colored solid curves show the RSS obtained by the proposed beam selection algorithm. The insets highlight the periodic beam refinement performed every 0.1~s. To evaluate the impact of beam nulling, UE-3 is intentionally nulled, and the purple dotted curve represents the received power at UE-3 after the beam-nulling operation. The received power at the nulled UE-3 remains  approximately between $-75$ and $-100$~dBm, approaching or falling below the $-97$ to $-88$~dBm thermal noise floor of 5G mmWave bandwidth over a substantial portion of the interval ~\cite{mmWaveBandWidth}. Despite nulling, other UEs maintain low RSS deviations, with power-gap RMSEs of $0.92$, $0.93$, and $0.95$~dB for UE-1, UE-2, and UE-4, respectively. 

\section{Conclusion}\label{sec:conclusion}

In this paper, we proposed a novel radar-assisted beam selection framework that exploits spatial sensing information to reduce beam management overhead. As a theoretical contribution, we derived the nearest-first selection overhead probability, providing a performance benchmark for related sensing-assisted beam selection schemes. In summary, we have provided a viable and low-overhead solution for beam management in practical UAV networks operating at mmWave frequency bands. Future work will focus on 
evaluating its performance in dynamic environments and validating its feasibility through hardware experiments. 

\bibliography{references}
\bibliographystyle{IEEEtran}
\end{document}